\documentclass[footinbib,aps,pra,reprint,superscriptaddress,nopacs]{revtex4-1}
\usepackage{amsmath,amssymb,bbm,bm,graphicx}
\usepackage[absolute]{textpos}
\usepackage{graphicx}
\usepackage{xcolor}

\DeclareMathOperator{\Tr}{Tr}

\begin{document}
\title{All-optical frequency processor for networking applications}

\author{Joseph M. Lukens}
\email{lukensjm@ornl.gov}
\affiliation{Quantum Information Science Group, Computational Sciences and Engineering Division, Oak Ridge National Laboratory, Oak Ridge, Tennessee 37831, USA}

\author{Hsuan-Hao Lu}
\affiliation{School of Electrical and Computer Engineering and Purdue Quantum Center, Purdue University, West Lafayette, Indiana 47907, USA}

\author{Bing Qi}
\affiliation{Quantum Information Science Group, Computational Sciences and Engineering Division, Oak Ridge National Laboratory, Oak Ridge, Tennessee 37831, USA}
\affiliation{Department of Physics and Astronomy, University of Tennessee, Knoxville, Tennessee 37996, USA}

\author{Pavel Lougovski}
\affiliation{Quantum Information Science Group, Computational Sciences and Engineering Division, Oak Ridge National Laboratory, Oak Ridge, Tennessee 37831, USA}

\author{Andrew M. Weiner}
\affiliation{School of Electrical and Computer Engineering and Purdue Quantum Center, Purdue University, West Lafayette, Indiana 47907, USA}

\author{Brian P. Williams}
\affiliation{Quantum Information Science Group, Computational Sciences and Engineering Division, Oak Ridge National Laboratory, Oak Ridge, Tennessee 37831, USA}
\date{\today}

\begin{abstract}
We propose an electro-optic approach for transparent optical networking, in which frequency channels are actively transformed into any desired mapping in a wavelength-multiplexed environment. Based on electro-optic phase modulators and Fourier-transform pulse shapers, our all-optical frequency processor (AFP) is examined numerically for the specific operations of frequency channel hopping and broadcasting, and found capable of implementing these transformations with favorable component requirements. Extending our analysis via a mutual-information--based metric for system optimization, we show how to optimize transformation performance under limited resources in a classical context, contrasting the results with those found using metrics motivated by quantum information, such as fidelity and success probability. Given its compatibility with on-chip implementation, as well as elimination of optical-to-electrical conversion in frequency channel switching, the AFP looks to offer valuable potential in silicon photonic network design.
\end{abstract}

\maketitle

\section{Introduction}
\label{intro}
Amid the persistent growth of data traffic and computational demands throughout the globe, photonic technologies are increasingly called upon to supplant electronic signal processing. Given the intrinsically high bandwidth available to optical carriers---coupled with low-loss transmission in optical fibers---photonics has successfully expanded into a variety of communications contexts, from long-haul spans to metro-area networks, datacenters~\cite{Agrell2016, Cheng2018}, and high-performance computing (HPC)~\cite{Taubenblatt2012, Rumley2015, Bergman2016, Miller2017}. As performance at the single-core level has plateaued in recent years, parallel architectures now lead the way in HPC, so that data \emph{movement} dominates much of the total power budget and thus can be viewed as perhaps the main hurdle on the path toward the Exascale regime ($10^{18}$ FLOPs/second)~\cite{Rumley2015}. Accordingly, the greater bandwidth and potentially much lower loss of optical technology---compared to the fundamental limits of copper wire~\cite{Miller1997, Miller2009}---have positioned optics as a leader in addressing these challenges, with CMOS-compatible silicon photonics offering unique potential in terms of both performance and scalability~\cite{Lim2014}.

It is therefore of utmost importance to design optical network topologies optimized for the needs of high-speed computing. Photonics' amenability to frequency parallelization proves invaluable in this objective. Each waveguide or fiber is capable of carrying channels covering many THz, and high-quality frequency-selective microring resonators (MRRs) are readily fabricated in silicon photonics; thus wavelength-division multiplexing (WDM) constitutes a natural, resource-effective foundation for photonic network architectures. Nonetheless, transferring data from one frequency channel onto another is a nontrivial task. From a practical side, the simplest approach is through optical-to-electrical-to-optical (OEO) conversion: detecting the symbols on one wavelength and electrically modulating them onto the desired output wavelength. However, such OEO conversion is unattractive, particularly for resource-intensive computing, for it introduces latency and dissipates extra energy.

\begin{figure*}[bt!]
\centering\includegraphics[width=1.3\columnwidth]{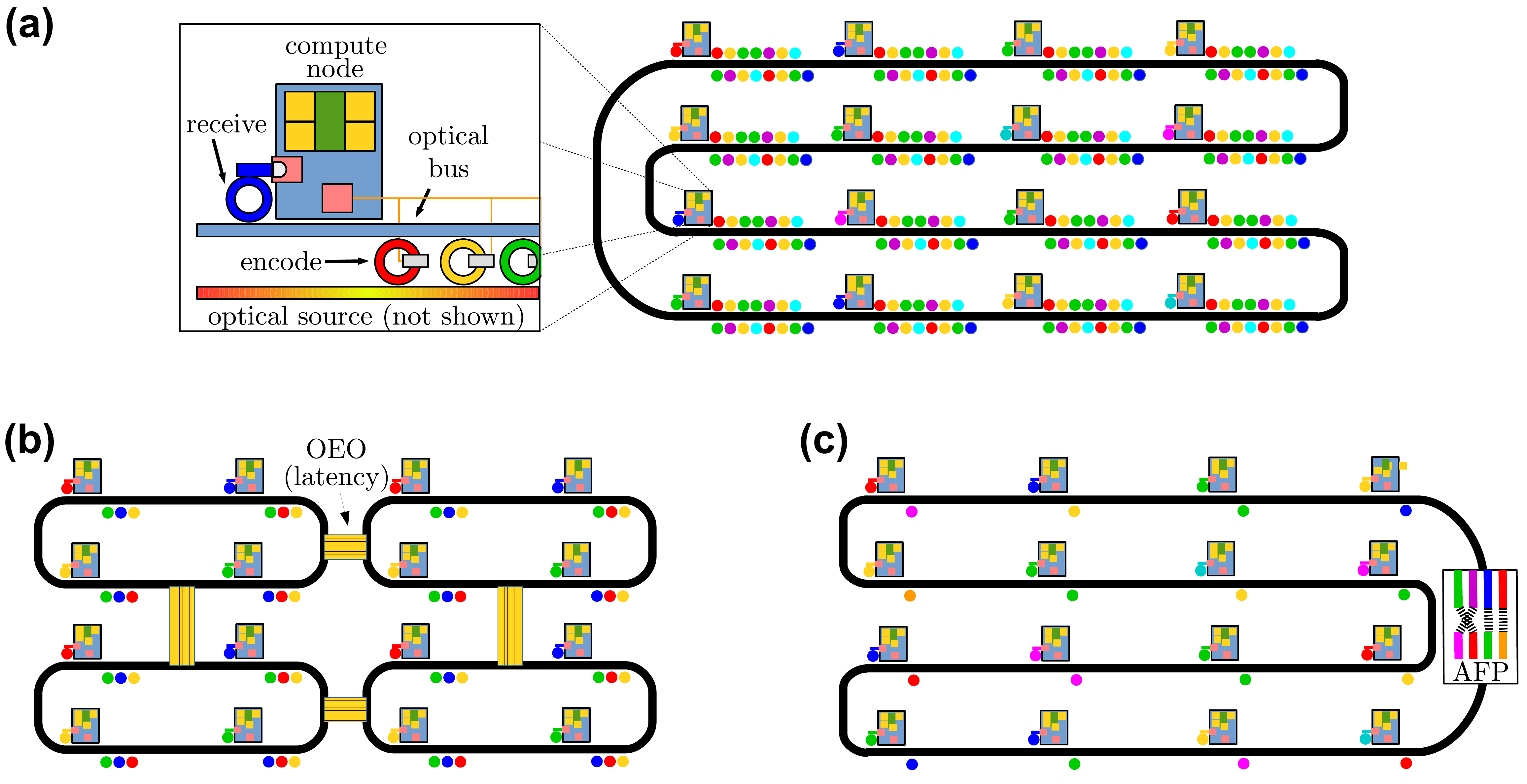}
\caption{\label{fig1} Comparison of optical network topologies. (a) Single bus waveguide, where each node communicates with all others by directly transmitting at the appropriate frequency. (b) Subnet design. Each node communicates optically with nodes inside its subnetwork, with OEO conversion used to send data between subnets. (c) Proposed design, in which an AFP transforms all input frequencies to outputs matched to each node's intended receiver.}
\end{figure*}

By contrast, optically transparent wavelength conversion can potentially eliminate these challenges, and has been the subject of long-standing research in the field of all-optical signal processing, where typically nonlinear optical interactions are recruited for ultrafast control and logic~\cite{Willner2014}. Recently, motivated by the application of quantum information processing~\cite{Nielsen2000}, we have introduced an alternative paradigm for all-optical wavelength control, termed the quantum frequency processor (QFP). Based on cascading electro-optic phase modulators (EOMs) and Fourier-transform pulse shapers, the QFP can in principle realize any unitary operation on frequency bins in a scalable fashion~\cite{Lukens2017a}, and several fundamental quantum gates have been demonstrated experimentally~\cite{Lu2018a, Lu2018b, Lu2019}. Importantly, the QFP's basic elements---modulators and frequency-selective phase shifters---are essential components of on-chip photonics, making the QFP an intriguing tool for future silicon photonic network design in \emph{classical} optics as well as quantum. In this Article, we propose and analyze a basic system for frequency multiplexing in which the same elements as a QFP function as an all-optical frequency processor (AFP): a centralized device to route traffic in a frequency-multiplexed network without OEO conversion. Using numerical optimization, we obtain designs for two basic network operations, frequency hopping and broadcasting. By then adopting an optimization metric based on mutual information, we also show how to construct systems for classical communication in resource-constrained environments. Overall, our results indicate significant potential in frequency-bin processing approaches for classical as well as quantum information manipulation, providing a general recipe for future designs optimized for specific networks.

This paper is organized as follows. In Sec.~\ref{concept} we introduce the basic idea of our AFP-based silicon photonic network, comparing it to alternative WDM configurations. Section~\ref{simulation} follows with details on our optimization model and simulation results for specific frequency transforms. We then present in Sec.~\ref{CC} an alternative optimization model able to incorporate practical limitations consistently using mutual information as the sole design metric. Finally, we discuss the implications of our results in Sec.~\ref{discussion} and conclude in Sec.~\ref{conclusion}.

\section{General Concept}
\label{concept}
Silicon photonic network designs vary widely in complexity, but typically include similar building blocks: sources, MRRs, modulators, and detectors~\cite{Rumley2015, Cheng2018}. For the purposes of our discussion here, a simple waveguide network of the form in Fig.~\ref{fig1}(a) is sufficient for the basic features of interest. Each computing node is designed to receive on one specific wavelength; for full connectivity, a filter bank containing $N$ modulators and MRRs is replicated at each node, with wavelengths matched to each of the other $N$ receivers. (The optical carriers can be generated independently at each node, or shared among many.) Since each node can communicate with any other by simply selecting the appropriate-wavelength MRR, network latency is low. However, the total number of MRRs grows quadratically with the number of users, increasing resource provisioning and total power burden. To improve scaling, one can partition into subnetworks, where inter-subnetwork nodes are accessed via OEO conversion at router interfaces [see Fig.~\ref{fig1}(b)]. While reducing provisioning, this approach suffers from increased latency within OEO operations. A variety of approaches for optimally balancing these demands have been discussed in the literature~\cite{Vantrease2008, Joshi2009, Pan2009,Li2014}, though the fundamental latency/resource tradeoff remains.

Motivated by this latency/resource conflict, we propose the network design in Fig.~\ref{fig1}(c). Here, each node has not only a unique receive wavelength, but also a fixed transmit wavelength, markedly reducing the resource requirements per node. The channel routing tasks are offloaded onto a centralized all-optical frequency processor (AFP) designed to actively convert the carrier frequencies of input data streams to match those of the desired destinations. Based on the QFP paradigm discussed above (a series of EOMs and pulse shapers), this AFP does include electro-optical manipulation controls, but it does not involve any form of OEO conversion: the frequency manipulations occur entirely in the optical domain, eliminating the associated latency and embodying the sense in which we apply the term ``all-optical.''

\begin{figure}[tb!]
\centering\includegraphics[width=\columnwidth]{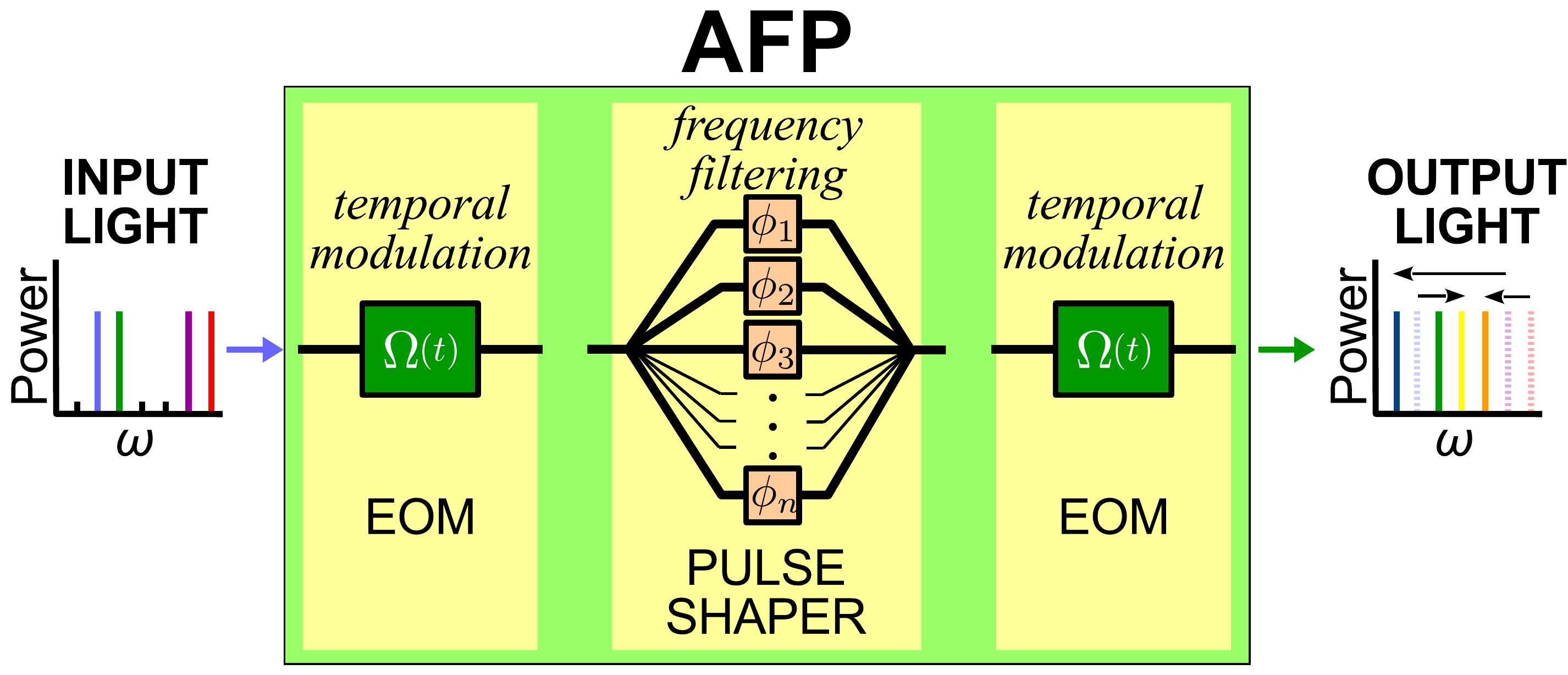}
\caption{\label{fig2} AFP example with $Q=3$ elements. Each element applies phase modulation in either time (EOM) or frequency (pulse shaper), realizing some desired frequency transformation.}
\end{figure}

A schematic of AFP construction follows in Fig.~\ref{fig2}. EOMs driven by radio-frequency (RF) waveforms periodic at the channel spacing are separated by pulse shapers that apply arbitrary phase shifts to each wavelength channel. A cascade of $Q$ such elements (defined as the sum of all EOMs and shapers) comprises the AFP. Theoretical considerations~\cite{Lukens2017a, Huhtanen2015} indicate that arbitrary unitary operations on $N$ modes are realizable with a number of elements $Q$ scaling no more rapidly than linearly with $N$, so that such a scheme should be able to realize any desired network connections with reasonable resource requirements. We should note, however, that the total number of MRRs is not necessarily smaller than in the original network of Fig.~\ref{fig1}(a); assuming an on-chip pulse shaper design with two MRRs per frequency mode~\cite{Wang2015}, and combining this with an anticipated linear scaling of the total number of pulse shapers with $N$, leads to a total growth in the number of MRRs proportional to $N^2$. Nevertheless, as we have discovered in the quantum regime~\cite{Lu2018a}, many operations can be realized with significantly fewer components than the theoretical limit, so it is important to focus on the specific connectivities desired in a particular network. And at the very least, the AFP can simplify node functionality demands by concentrating the more difficult network arbitration capabilities at a central location.

To move beyond the high-level theoretical characteristics of our proposal, for the remainder of this Article we concentrate on explicit designs for AFP configurations performing basic channel routing tasks. The variety of node connections needed for all network operating states is far too vast for us to consider all possibilities directly, though we see two as particularly foundational and instructive: a cyclic frequency hop and 1-to-$N$ channel broadcast. Consider a single-spatio/polarization-mode optical field on the bus waveguide [$E(t)=E^{(+)}(t) + \mathrm{c.c.}$], with
\begin{equation}
\label{e1}
E^{(+)}(t) = \sum_{n=0}^{N-1} a_n(t) e^{-i\omega_n t},
\end{equation}
where the individual carriers are assumed equispaced with separation $\Delta\omega$ ($\omega_n = \omega_0 + n\Delta\omega$), and the baseband data streams $a_n(t)$ are slowly varying with respect to $1/\Delta\omega$, which ensures frequency manipulation is possible without introducing distortion in the signals; in other words, the data streams can be transformed on a symbol-by-symbol basis. The AFP operation can then be modeled as a matrix $V_{nm}$ transforming inputs $a_n(t)$ to outputs $b_n(t$) according to:
\begin{equation}
\label{e2}
b_m(t) = \sum_{n=0}^{N-1} V_{mn}a_n(t).
\end{equation}
The matrix $V_{mn}$ is related to the underlying pulse shaper and EOM transformations identically as in the fully quantum analysis in Ref.~\cite{Lukens2017a}; the only difference now is that the fields are treated classically, rather than as bosonic operators. Additionally we note that $V$ is unitary when considered over all possible frequency bins, that is,
\begin{equation}
\label{e2-1}
\sum_{n=-\infty}^\infty V_{nm}^* V_{np} = \delta[m-p],
\end{equation}
where $\delta[n]$ is the Kronecker delta function. Insertion loss in practice introduces an overall scaling factor; but this effect is absent from ideal phase modulation, so we do not consider it in the initial design simulations. 

For the fundamental frequency-hop operation, we consider the $N\times N$ permutation matrix $S_N$, with elements
\begin{equation}
\label{e3}
(S_N)_{mn} = \delta[(m-n-1)\,\textrm{mod}\,N],
\end{equation}
and $m,n \in \{0,1,...,N-1\}$. In words, this transformation hops the fields at each frequency according to the prescription $\omega_0 \rightarrow \omega_1, \omega_1 \rightarrow \omega_2,...,\omega_{N-1} \rightarrow \omega_0$. All other possible shifts that preserve this sense of ordering can then be written as powers: $S_N, S_N^2,...,S_N^{N-1}$. ($S_N^N$ returns the identity and the sequence repeats.) Of these $N-1$, only powers through $\mathrm{floor}(N/2)$ need be considered in design, as the remaining are simply transposes which can be obtained physically by reversing element order and conjugating all phases. While this permutation set does not encompass all possible one-to-one frequency channel configurations, it does include all $N$ hops for a specific channel, while simultaneously demanding hopping over the remaining $N-1$. In this sense, the configurations are more demanding than, e.g., asking only one channel to hop while keeping the remaining channels fixed.

The second major capability we consider for our AFP design is a broadcast transformation, which copies the data stream from one channel onto all $N$ frequencies. One unitary transformation which accomplishes this is the $N$-point discrete Fourier transform (DFT), whose elements are
\begin{equation}
\label{e4}
(F_N)_{mn} = \frac{1}{\sqrt{N}} e^{2\pi i \frac{mn}{N}}.
\end{equation}
Again the indices are defined such that $m,n \in \{0,1,...,N-1\}$. This operation spreads the data stream from one of the input channels to all $N$ wavelengths, with a corresponding reduction in power of $1/N$ (satisfying energy conservation). Practically speaking, such a broadcast is valuable when the other input channels are either quiet or modulating on orthogonal temporal modes; otherwise, the symbols will interfere at the output. As with the permutation matrices, one could envision less-demanding broadcast configurations---optimized for only one of the $N$ channels to cast, not all of them---so we can view $F_N$ as a universal broadcast for a particular $N$-node network.

\section{Numerical Optimization}
\label{simulation}
With these two characteristic operations defined, our main objective centers on how to efficiently implement them in hardware. We enlist the numerical procedure introduced in~\cite{Lukens2017a} for determining optimal solutions. In the slowly varying symbol regime [c.f. Eq.~(\ref{e1}) and following], the system operation can be viewed as fully discrete in the frequency domain, and thus all EOM patterns repeat with period $2\pi/\Delta\omega$ set by the frequency-bin spacing. For simulation purposes, we discretize this temporal period into $M$ samples and truncate the number of frequency modes to $M$ as well. Given the possibility of input channels scattering outside of the desired $N$-channel space, we choose $M$ such that $M\gg N$. In other words, $M$ must be sufficiently large to fully encompass all bins occupied by frequency channels throughout an operation.

In the $M$-bin discretization, each pulse shaper acts as an $M\times M$ diagonal unitary $D$ over frequency bins, multiplying each frequency channel by some arbitrary phase.  Conversely, an EOM is represented as an $M\times M$ diagonal unitary $D$ operating on \emph{time} samples, or as $F_M D F_M^\dagger$ in the frequency domain, where we approximate the Fourier transform according to the DFT matrix $F_M$ [cf. Eq.~(\ref{e4})] defined over $M$ samples. Then a full network of $Q$ elements becomes
\begin{equation}
\label{e5}
V = F_M D_Q F_M^\dagger \cdots D_2 F_M D_1 F_M^\dagger,
\end{equation}
where we have assumed the first element is an EOM and $Q$ is odd. It is important to note that the DFT matrices in this expression serve as a numerical approximation for the continuous temporal space, in contrast to the $N$-point DFT considered as the fundamental broadcast operation, which is not an approximation but rather the exact transformation we seek to realize on a dimension-$N$ subset of physical frequency modes. Additionally, throughout this discretization process, we preserve unitarity by specification, as $V$ is formed by a product of fully unitary matrices.

To assess the quality of a particular AFP configuration [Eq.~(\ref{e5})] relative to the desired $N\times N$ transformation $T$ (frequency hop or broadcast), we first define $W$ as the $N\times N$ submatrix of $V$ within the channel modes of interest. This $W$ fully characterizes the operation when considering $N$ channels in and the same $N$ out, and it may or may not prove unitary, depending on the specific transformation. As long as $M$ is sufficiently large so that the numerical approximation of $V$ [Eq.~(\ref{e5})] is valid in the $N\times N$ projection, $W$ will correspond to the true experimentally realizable operation. We then classify the performance of $W$ compared to $T$ according to fidelity
\begin{equation}
\label{e6}
\mathcal{F} = \frac{ \Tr (W^\dagger T) \Tr (T^\dagger W) }{\Tr (W^\dagger W)  \Tr (T^\dagger T)}
\end{equation}
and success probability
\begin{equation}
\label{e7}
\mathcal{P} = \frac{\Tr (W^\dagger W)}{\Tr ( T^\dagger T)}.
\end{equation}
These are identical to the metrics previously considered in the context of quantum information processing~\cite{Lukens2017a, Uskov2009}; the condition $\mathcal{F}=\mathcal{P}=1$ signifies the situation $W=e^{i\phi} T$, with $\phi$ an unimportant global phase. In the case of classical optical communications, the quantum fidelity $\mathcal{F}$ is related to the purity of the operation, while $\mathcal{P}$ quantifies the overall efficiency. Since the EOM/pulse shaper operations are modeled as unitary matrices, such loss ($\mathcal{P}<1$) corresponds to power scattering outside of the $N$-channel subspace into adjacent frequency bins, rather than to actual photon absorption.

Finally, before diving into the numerical results, we note that the chosen frequency-hopping [Eq.~(\ref{e3})] and broadcast [Eq.~(\ref{e4})] matrices are themselves related by a straightforward decomposition: $S_N^n = F_N^\dagger D_N^n F_N$, where $D_N$ is a diagonal matrix consisting of all $N$th roots of unity, i.e., $(D_N)_{mm} = e^{2\pi i m/N}$. This relationship implies that if one can realize the $N$-channel DFT, a permutation of any power $n$ follows simply by adding a pulse shaper ($D_N^n$) and a second (conjugated) DFT. While interesting on a formal level, such a construction requires more than double the number of elements compared to the given solution for $F_N$, undesirable from a resource perspective; thus, in the following simulations we look to synthesize permutations directly, rather than building on broadcast solutions. The two matrix classes likewise find connections in quantum information as well. Indeed, the $N=2$ incarnations, $S_2$ and $F_2$, represent the Pauli-$X$ and Hadamard gates, respectively, and both $F_2$ and $F_3$ have been experimentally realized in the QFP paradigm~\cite{Lu2018a}. On the other hand, no example of an $S_N$ has been shown with a QFP. Incidentally, this distinction in progress between the permutation and DFT gates follows from differences in their functionality. For whereas frequency hopping demands accurate concentration of an input frequency channel into \emph{one} specific output mode, with minimal leakage into the remaining channels, broadcasting seeks the opposite: spread a channel's information into \emph{all} output modes. Accordingly, in this sense these two operations occupy extremes in the wider class of useful AFP transformations, making them extremely valuable test cases.

\subsection{Arbitrary Modulation}
In the first round of simulations, we consider the case in which each EOM is permitted arbitrary modulation patterns; i.e., all $M$ elements within a temporal period can assume any value in the interval $(-\pi,\pi]$. In order to ensure the solution does not contain numerical artifacts from reaching the edge of the mode space and wrapping around in an unphysical manner (i.e., aliasing), we limit the number of nonzero pulse shaper phases to 32, one quarter of the total number of bins considered in the full discretized space, $M=128$. For the optimization procedure itself, we constrain $\mathcal{F}\geq0.99$ and seek a set of phase values which maximize $\mathcal{P}$. One could alternatively constrain $\mathcal{P}$, or maximize to the product $\mathcal{FP}$; for these first tests, we focus on the $\mathcal{F}$ constraint so that every solution will perform the desired operation well, only with a reduction in power specified by $\mathcal{P}$. We have found that an odd number of elements $Q$, in which EOMs comprise the first and last devices in the series, is generally more powerful than an even number; that is, adding a pulse shaper on the front or back end has not been seen to boost performance (with the sole exception of the operation $F_2$). Therefore we consider AFPs consisting of $Q\in\{1,3,5,...\}$ components. Making use of nonlinear optimization in MATLAB, we run multiple optimizations with either random values or previous solutions as starting points, and report the final solution with maximum $\mathcal{P}$ satisfying $\mathcal{F}\geq 0.99$.

\begin{figure}[t!]
\centering\includegraphics[width=\columnwidth]{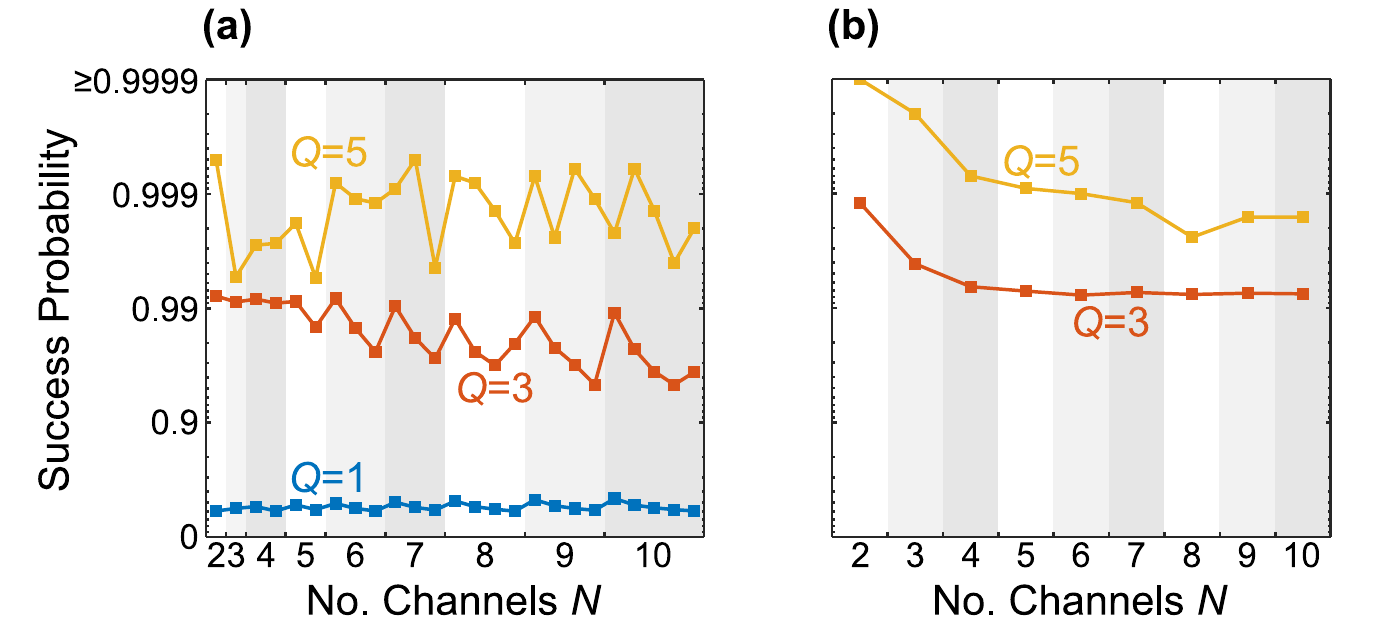}
\caption{\label{fig3} Optimal solutions for arbitrary temporal modulation patterns, with $\mathcal{F}\geq 0.99$ as constraint. (a) Frequency hopping matrices $S_N^n$. (b) Broadcast (DFT) matrices $F_N$. In both plots, vertical shading divides the results by dimension $N$. In (a), the solutions for a given $N$ are ordered left-to-right by increasing power $n\in\{1,2,...,\mathrm{floor}(N/2)\}$.
}
\end{figure}

Figure~\ref{fig3}(a) presents our findings for all unique permutations (frequency hops) for $N=2$ to $N=10$ channels, a total of 25 configurations. The points for each $N$ are sorted left-to-right by power $S_N^n$ [$n\in\{1,2,...,\mathrm{floor}(N/2)\}$]. For $Q=3$, the shorter hops (smaller $n$) tend to perform better in terms of success $\mathcal{P}$; for $Q=5$, the number of elements is sufficiently high for roughly uniform performance with $n$ (note the logarithmic vertical axis). In total, $Q=3$ is able to realize all 25 transformations with $\mathcal{P}>0.95$, and $Q=5$ with $\mathcal{P}>0.994$. While moving significantly beyond $N=10$ is prohibited by our current computational capabilities, these results certainly suggest the possibility of extremely favorable \emph{sublinear} scaling of the number of components with network size $N$. The simulation findings for the DFT broadcast operation follow in Fig.~\ref{fig3}(b), also for $N=2$ to $N=10$ channels. In this case, a single EOM is unable to satisfy the constraint $\mathcal{F}\geq 0.99$, so $Q=1$ has no solutions in this plot. However, the $Q=3$ solutions perform even better than for frequency hopping; $\mathcal{P}>0.99$ for all configurations examined, and increasing to $Q=5$ boosts these values closer to unity. 

Comparing the results for both $S_N$ and $F_N$, we see that the frequency hops summarized in Fig.~\ref{fig3}(a) possess relatively jagged scaling curves, likely due to differences in the complexity of permutations that span different numbers of channels (powers of $S_N$). Yet nonetheless, taken as a whole, the success probabilities for both frequency hopping and broadcasting are fairly flat, not decaying rapidly with increasing $N$. Such behavior for fixed and relatively small AFP sizes ($Q\leq 5$) provides preliminary evidence of improved network scaling within the AFP paradigm. Nevertheless, further simulations for larger networks---and perhaps additional transformations beyond $S_N$ and $F_N$ primitives---will be required to pin down precisely the resource requirements for an AFP in specific applications.



\subsection{Sinewave Modulation}
While the fully arbitrary modulation patterns in the previous section probe the most general capabilities of AFPs, complex waveforms can prove difficult to implement experimentally, particularly at common frequency channel spacings. For example, at 25~GHz channel separation (a typical value in dense WDM), an industry-leading 120~GS/s arbitrary waveform generator~\cite{Keysight2019} provides only 4.8 points per period at the fundamental tone. In contrast, high-frequency sinewave RF drives are readily obtained from analog generators, making AFP configurations based on the simpler case of single-frequency electro-optic modulation a worthwhile subset to examine for practical purposes. To do so, we next restrict the phase applied by each EOM to a sinewave with to-be-determined amplitude and phase, otherwise keeping the optimization procedure identical to above.

\begin{figure}[tb!]
\centering\includegraphics[width=\columnwidth]{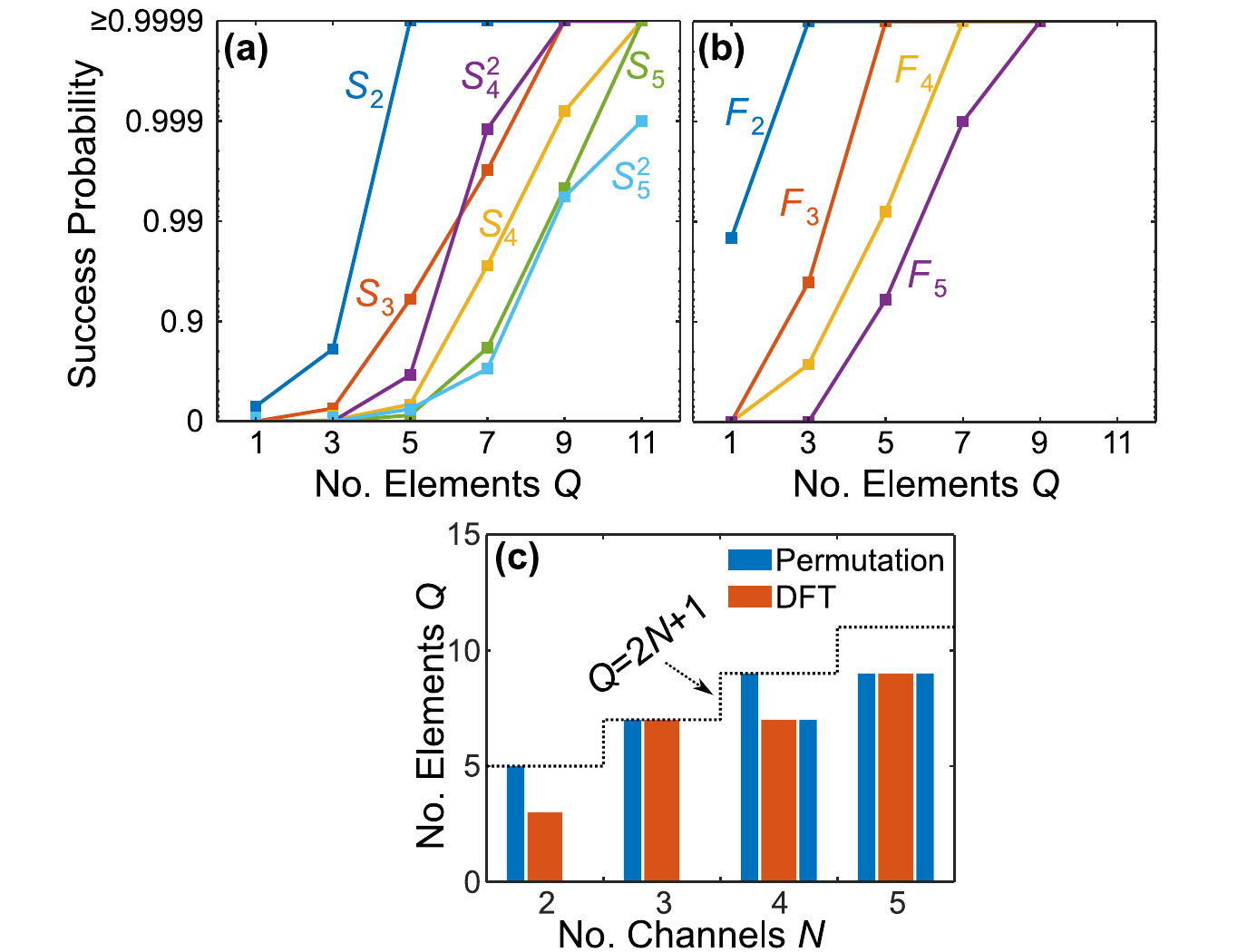}
\caption{\label{fig4} Simulation results under the restriction of sinewave-only temporal modulation. Success probabilities for (a) permutations $S_N^n$ and (b) DFTs $F_N$, plotted against the number of AFP elements. (c) Number of elements needed to reach $\mathcal{P}\geq0.99$ for the transformations in (a) and (b).}
\end{figure}

The maximal success probabilities for the hopping operations (with $\mathcal{F}\geq 0.99$) are shown in Fig.~\ref{fig4}(a); the broadcast solutions follow in Fig.~\ref{fig4}(b). Due to the increase in $Q$ depth---and concomitant demands on computational resources---we simulate only through $N=5$ channels here. With fewer matrices, but more AFP sizes, we now plot each transformation as a separate curve and take $Q$ as the abscissa. Unlike the arbitrary modulation findings, the number of components required for a given success $\mathcal{P}$ does increase strongly with matrix size $N$. To explore this observation more quantitatively, we consider the minimum number of elements $Q$ required for each transformation to reach the threshold $\mathcal{P}\geq 0.99$, as summarized in Fig.~\ref{fig4}(c). Both permutation and DFT results are binned by dimension $N$ ($N=4,5$ each have two distinct permutations shown). While difficult to extrapolate with such a limited number of data points, we observe that all matrices fall under the scaling cap $Q=2N+1$; in other words, for a given $N$, no operation requires more than $2N+1$ elements to be realized with $\mathcal{F}, \mathcal{P} \geq 0.99$. Should this cap hold for larger dimensions as well, it would show that---at least for these particular matrices---the linear scaling $Q \propto N$ expected for \emph{arbitrary} temporal/spectral AFP patterns~\cite{Lukens2017a, Huhtanen2015} can hold even for the significantly restricted class of sinewave-only temporal modulation.

In order to examine scaling in our AFP approach in greater detail, though, it is important not only consider the circuit depth $Q$, but also the effective bandwidth. A large portion of the resource requirements for a silicon photonic network rests on the total number of MRRs; and because some scattering into frequency channels outside of the $N$-mode network occurs in the AFP, it is valuable to quantify precisely how many wavelengths must be individually addressed by each pulse shaper. Unfortunately, given the relatively few matrices we have successfully simulated, our initial attempts to determine the scaling of the number of MRRs with channels $N$ remain inconclusive. It seems possible that linear resource scaling may hold, in which case the AFP will eventually win out in terms of resources compared to networks modeled after Fig.~\ref{fig1}(a). Developing heuristic models which could allow extrapolation to much larger dimensions thus forms an important objective for future studies.

Finally, we mention one more EOM synthesis approach falling between the two extremes of arbitrary waveforms and single sinewaves: RF patterns consisting of the fundamental tone plus low-order harmonics. For example, an EOM operating on 25~GHz channels could utilize a sum of 25, 50, and 75~GHz components for more precise control of the optical sidebands. Given the availability of frequency multipliers reaching into the W-band (75--110 GHz)---as well as possibilities to generate multitone RF signals directly using line-by-line optical pulse shaping and fast photodetection~\cite{Huang2007,Victor2012,Victor2013}---such Fourier series-like signal construction may prove more feasible than direct digital synthesis with arbitrary waveform generators. To test possibilities enabled by such multiharmonic signals, we have repeated the sinewave-only optimization, adding harmonics with adjustable amplitude and phase. For the frequency hopping transformations, our simulations have suggested that adding RF harmonics can indeed improve success $\mathcal{P}$ for a given $Q$. In the case of the broadcast (DFT) operation in particular, we have noticed fascinating behavior when increasing the number of harmonics. Specifically, we have found solutions for all DFT matrices from $N=2$ to $N=5$ maintaining $\mathcal{F}\geq 0.99$ and $\mathcal{P}\geq 0.98$, using only $Q=3$ elements, but exploiting EOM signals with a total of $N-1$ tones: e.g., just the fundamental for $N=2$, fundamental plus next harmonic for $N=3$, etc. Thus in this particular case, harmonic addition seems to provide a viable alternative to cascading additional components. Indeed, this approach was already demonstrated for the case $N=3$, $Q=3$, in a frequency-bin tritter~\cite{Lu2018a}. The additional simulations here suggest that, rather than a coincidence for the tritter, harmonic addition may in fact be a general design feature of the frequency-bin DFT.

\section{Optimization for Limited Resources}
\label{CC}
When considering frequency-bin operations in either a classical (AFP) or quantum (QFP) context, any design attaining both $\mathcal{F}\rightarrow 1$ and $\mathcal{P}\rightarrow 1$ is fully optimal, in that it replicates the desired transformation perfectly. On the other hand, for configurations which fail to reach unity on one or both of $\mathcal{F}$, $\mathcal{P}$---e.g., due to physical limits on the number of elements or complexity of temporal modulation---it is possible that the optimal solution for a quantum gate may prove suboptimal for the equivalent classical AFP transformation. For quantum applications such as discrete-variable QIP, $\mathcal{F}$ and $\mathcal{P}$ possess well-defined and operationally significant meanings: success $\mathcal{P}$ denotes the probability that the gate will succeed, as defined by photons exiting the setup within a predefined set of modes, while $\mathcal{F}$ quantifies a successful operation's closeness to the ideal manipulation. Because of the binary nature of the multiphoton output---i.e., the photons are either in the desired output channels, or not---such a clear distinction between $\mathcal{F}$ and $\mathcal{P}$ appears naturally. By contrast, in classical signal processing~\cite{footnote}, information resides in the macroscopic properties of an optical field, such as amplitude and phase. Therefore, faithful data transmission rests on being able to distinguish between the outputs corresponding to each of the input symbols, an objective impacted both by total signal amplitude (related to $\mathcal{P}$) and by additional noise (related to $\mathcal{F}$). For this reason, when the ideal $\mathcal{F}=\mathcal{P}=1$ is unattainable in practice, it is unclear \emph{a priori} which fidelity/probability tradeoff is optimal.

To answer this question, we consider a new optimization metric, mutual information, which expresses the number of bits shared between two random variables~\cite{Shannon1948, Cover2006}, in our case the sent and received symbols. Mutual information aggregates all design metrics into a single number, removing any question of artificial balancing of potentially conflicting demands. The price for doing so, however, is the need to specialize to a particular encoding format, noise model, and power level (energy per symbol). Whereas $\mathcal{F}$ and $\mathcal{P}$ are universal in the sense that they depend only on the AFP itself, irrespective of the input signal and noise properties, mutual information is directly affected by these considerations; the optimal transformation may depend profoundly on, e.g., whether the input signal-to-noise ratio (SNR) or interchannel crosstalk dominates errors.

We make these initial observations concrete by specifying just such a model. We consider $N$ input frequency channels, each carrying equal average power and independent, Gaussian modulated data---known to attain the channel capacity bound for a given SNR under additive white Gaussian noise (AWGN)~\cite{Shannon1948, Cover2006, Giovannetti2014}. Assuming Gaussian modulation with zero mean and variance $\sigma^2$ in both available quadratures ($X$ and $Y$), the average photon number per complex symbol satisfies $\mu=2\sigma^2$ in our normalization. At the output, we assume conjugate homodyne detection with an optical hybrid, so that both $X$ and $Y$ quadratures are measured simultaneously at each frequency.

In a single output quadrature, the $N\times N$ channel matrix $W$ produces a variance in channel $k$ from input $l$ given by $K^2\mu_\mathrm{LO}\eta\sigma^2|W_{kl}|^2$, where $K$ is an optical-to-electrical system conversion factor (e.g., photons/pulse to symbol in volts), $\mu_\mathrm{LO}$ the number of photons per symbol contained in the local oscillator (LO) entering the hybrid, and $\eta$ is an efficiency parameter encompassing all transmissivities in the system extrinsic to the AFP operation (e.g., detector and component insertion losses), which we take as equal for all input channels. Because of rotational symmetry in dual-quadrature Gaussian modulation, each input frequency $\omega_n$ will contribute crosstalk noise which is also Gaussian of variance $K^2\mu_\mathrm{LO}\eta\sigma^2|W_{kn}|^2$, regardless of the phase relationship between various input channels (drifting randomly over $2\pi$ or permanently fixed). Finally, we model the detection noise as Gaussian with variance $(1+D)K^2\frac{\mu_\mathrm{LO}}{2}$: the first term corresponds to the vacuum noise in each quadrature, and $D$ represents additional electronic noise normalized by the vacuum level. Since $\mu_\mathrm{LO}\gg \eta\mu$ in this model, we can safely neglect any modifications to the variance from the signal itself.

Accordingly, the SNR for either the $X$ or $Y$ quadrature for the $kl$ connection becomes
\begin{equation}
\label{e7-1}
R_{kl} = \frac{\eta\mu |W_{kl}|^2}{1+D+\eta\mu \sum_{n\neq l} |W_{kn}|^2},
\end{equation}
where we express the modulation variance in terms of average photon number per symbol ($\mu=2\sigma^2$). Because the total noise, including crosstalk, is still AWGN, the mutual information between channel $k$ at the output and $l$ at the input attains the channel capacity, $\frac{1}{2}\log_2(1+R_{kl})$ per quadrature, or
\begin{equation}
\label{e8}
I_{kl} = \log_2 \left[ \frac{1 + \mu_\mathrm{eff} P_k}{1 + \mu_\mathrm{eff} P_k (1-C_{kl})} \right]
\end{equation}
in total (summing $X$ and $Y$). Here we have introduced an effective photon number $\mu_\mathrm{eff} = \frac{\eta\mu}{1+D}$, that rescales $\mu$ by the non-AFP noise sources, and defined the channel probability
\begin{equation}
\label{e9}
P_k = \sum_{n=0}^{N-1} |W_{kn}|^2,
\end{equation}
which falls in the interval $[0,1]$ because of the unitarity of the matrix $V$ from which $W$ is derived. We also have specified the selectivity as
\begin{equation}
\label{e10}
C_{kl} = \frac{|W_{kl}|^2}{P_k}.
\end{equation}
Intuitively, we can begin to see useful relationships between these metrics and the previous $\mathcal{F}$ and $\mathcal{P}$. Indeed, $\mathcal{P} = \frac{1}{N} \sum_{k=0}^{N-1} P_k$ directly; it is nothing but the average throughput for each channel.  Similarly, the selectivity $C_{kl}$ is highly related to fidelity $\mathcal{F}$, though not in such explicit terms. In particular, $C_{kl}$ (and the mutual information more generally) has no dependence on the phase of the elements in $W$, as a consequence of the fact each frequency channel is a separate, independent data carrier. On the other hand, $\mathcal{P}=1$ if and only if $P_k=1 \,\forall\,k$. Finally, we see that the limiting case of $P_k=C_{kl}=1$ gives $I_{kl}=\log_2(1+\mu_\mathrm{eff})$, as expected for perfect dual-quadrature Gaussian modulation.

The above formula [Eq.~(\ref{e8})], which includes crosstalk effects from adjacent co-transmitting frequencies, aligns well with the scenario of frequency hopping under its brightest operating condition (all channels communicating). On the other hand, broadcasting to all outputs is meaningful only when one input alone is transmitting. 
And so for this case, we modify the mutual information formula by removing noise effects resulting from interchannel crosstalk, leaving the simpler expression
\begin{equation}
\label{e11}
I_{kl} = \log_2 \left( 1 + \mu_\mathrm{eff} P_k C_{kl} \right).
\end{equation}
For example, the case of the ideal DFT operation [Eq.~(\ref{e4})] has $P_k=1, C_{kl}=\frac{1}{N}\,\forall \,k,l$, implying a $N^{-1}$ reduction in SNR for each output compared to the input, by energy conservation. In the frequency-hopping scenario, a total of $N$ mutual informations $I_{kl}$ prove important: $k\in\{0,1,...,N-1\}$ and $l=f(k)$, where $f(k)$ is the one-to-one function specifying the desired I/O connections. Since the broadcast configuration enables any of the $N$ inputs to cast (though not simultaneously), all $N^2$ mutual information pairs expressed by Eq.~(\ref{e11}) should be accounted for in the solver. For either situation, we have found that choosing the average of all relevant I/O paths as optimization metric tends to produce solutions with widely varying performance across all channels. So to improve uniformity, we select the minimum value over all channels as the metric. In other words, while each specific set of AFP settings produces several mutual information values $I_{kl}$ of interest, the optimizer only considers the smallest one---regardless of $k,l$---in rating the quality of that particular configuration. This procedure tends to improve channel homogeneity by effectively redirecting parameter resources toward whichever channel lags the others in the current iteration of the optimizer.

\begin{figure}[tb!]
\centering\includegraphics[width=1\columnwidth]{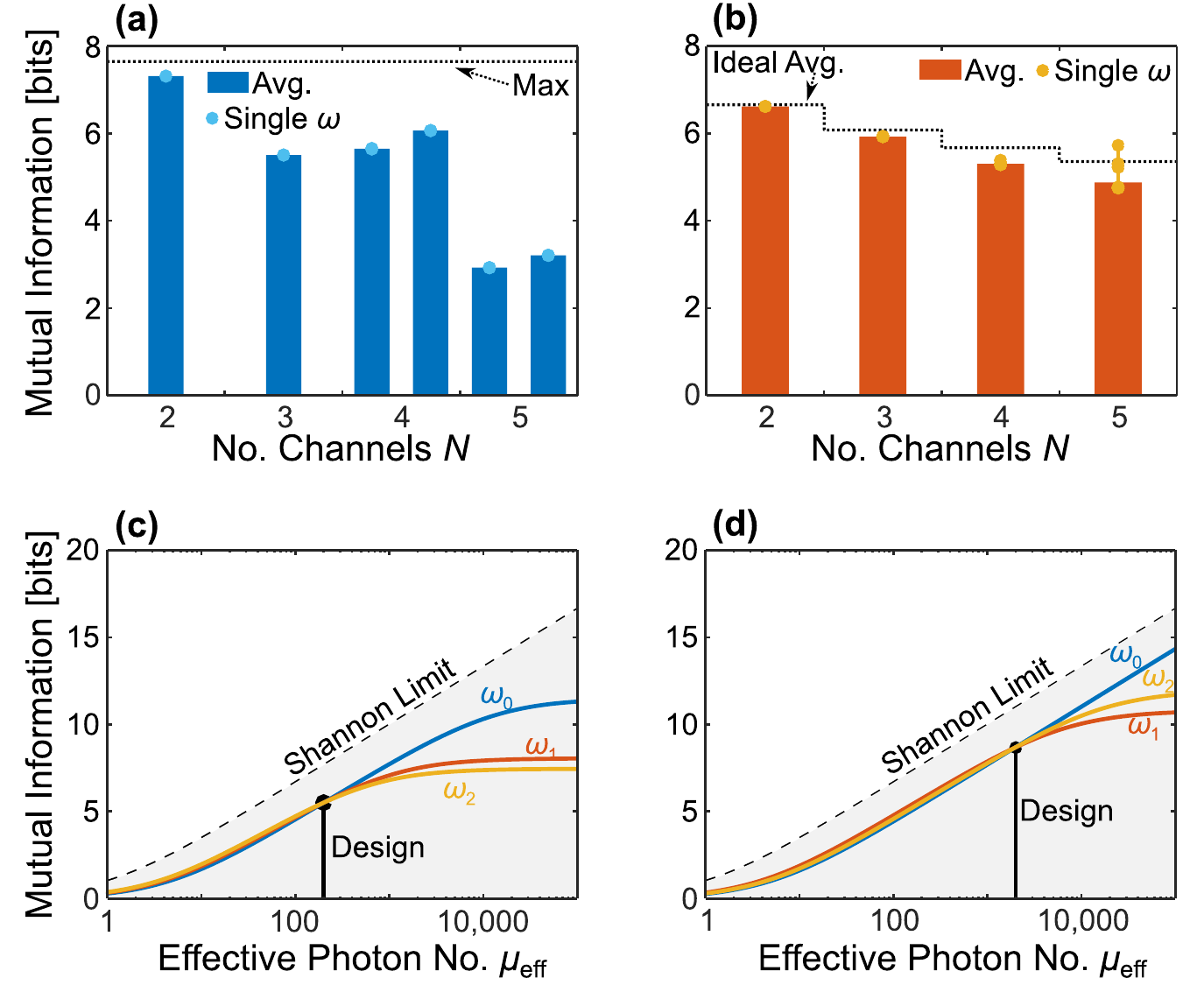}
\caption{\label{fig5} Mutual information simulations for $Q=3$ elements and sinewave-only modulation. Findings for (a) frequency hopping operations through $N=5$ and (b) $1$-to-$N$ broadcasting. (c) Dependence of individual mutual information values on photon number $\mu_\mathrm{eff}$ for the $N=3$ solution in (a). (d) Scaling for a second hopping solution optimized for $\mu_\mathrm{eff}=2000$. Each curve in (c) and (d) is labeled by the carrier frequency of the output channel, $\omega_k$, where $k\in\{0,1,2\}$.
}
\end{figure}

Because the focus on mutual-information--based metrics is motivated by the practical constraints encountered in a real-world system, we intentionally limit the number of elements to $Q=3$ and the available modulation to sinewave-only, exploring designs for $N\in\{2,3,4,5\}$ channels in numerical optimization. As evident in both Eqs.~(\ref{e8}) and (\ref{e11}), one must now specify the effective photon number $\mu_\mathrm{eff}$ at the onset. Figure~\ref{fig5}(a) plots the mutual information results obtained for the hopping operation with $Q=3$ elements and $\mu_\mathrm{eff}=200$. The bars signify averages over all $N$ hops for a particular matrix, while the separate values for each channel are represented by lighter dots (which are too close to be distinguished on this scale). As before, $N=4$ and $N=5$ have two distinct transformations represented. At this photon number, the Shannon-limited channel capacity (perfect hopping) is 7.65 bits (dotted line); the setup gets extremely close to this limit for $N=2$, trailing by wider gaps as the number of channels increases, which makes sense given the fixed number of resources. 

The results for the broadcast operation are plotted in Fig.~\ref{fig5}(b). As in the hopping cases, the solid bars mark the average for each solution, and the lighter dots show the individual channel results. Here the ideal mutual information (dotted line) now varies with $N$, as a consequence of sharing photonic energy among $N$ modes. And because it is an average, unlike the Shannon limit in the hopping simulations, individual channels can exceed it, which does occur in the $N=5$ solution. Overall, the solutions perform better than than the mode hopping tests, with all cases within 10\% of the ideal average. Incidentally, the high mutual information in the $N=5$ case contrasts markedly with the DFT simulations of Fig.~\ref{fig4}(b), where no solution with nonzero success probability was obtained for the matrix $F_5$ with $Q=3$ elements and $\mathcal{F}>0.99$. Unlike the expression for fidelity [Eq.~(\ref{e6})], in Eq.~(\ref{e11}) the phases of the $W_{kn}$ matrix elements do not appear, thereby significantly relaxing demands on the transformations. This situation illuminates the importance of mutual information in streamlining system design requirements.

These solutions offer valuable insights into the interplay between channel probability [Eq.~(\ref{e9})] and selectivity [Eq.~(\ref{e10})]. For example, the squared-moduli of the elements for the $N=3$ channel hop solution are
\begin{equation}
\label{e12}
|W|^2 =
\begin{pmatrix}
0.00003 & 0.00005 & 0.22594 \\
0.26664 & 0.00094 & 0.00005 \\
0.00110 & 0.29725 & 0.00058
\end{pmatrix}
\end{equation}
Although all channels attain near-identical mutual information at $\mu_\mathrm{eff}=200$, their respective $P_k$ and $C_{kl}$ values vary. For example, the $\omega_0$ output (top row) has appreciably lower probability, but higher selectivity, compared to $\omega_1$ and $\omega_2$. Evidently, the extra flexibility available to the system in letting these values vary across channels enables higher mutual information than that possible by requiring completely uniform amplitudes. The effect of such interchannel differences becomes clear when examining this solution at photon numbers other than the designed value. In Fig.~\ref{fig5}(c), we plot the mutual information computed from the above matrix for $\mu_\mathrm{eff}$ ranging from 1 to $10^{5}$. Above the designed value of $\mu_\mathrm{eff}=200$, channel $\omega_0$ performs best; with the extra power available, the data rate is primarily limited by crosstalk from other channels, so $\omega_0$'s high selectivity grants it an edge. On the other hand, the situation reverses at lower photon numbers, where simply receiving enough photons has the greatest impact; for $\mu_\mathrm{eff}<200$, $\omega_0$'s lower probability places it at the bottom of the three channels. As one more example, we plot in Fig.~\ref{fig5}(d) the mutual information values for a solution optimizing the same $N=3$ channel hop, but for $\mu_\mathrm{eff}=2000$. Now all three channels coincide at this higher photon level, with variations moving away from it, again reflecting the specific matrix elements. Finally, we can also directly compare the optimal solutions at different design values of $\mu_\mathrm{eff}$, where we see a similar trend, in which the optimizer tends to find solutions with better selectivity given higher photon counts. Considering the two design values in Fig.~\ref{fig5}(c) and (d), the average selectivity over all channels improves from $0.9968$ to $0.9997$ when $\mu_\mathrm{eff}$ is increased from $200$ to $2000$, and correspondingly the average success probability drops from $0.26$ to $0.23$. Accordingly, these simulations highlight how the mutual information approach balances the parallel pulls of low crosstalk and high overall throughput, as well as the importance of the noise model and power level in resource-limited network designs.

\section{Discussion}
\label{discussion}
The research problem we have undertaken here---application of frequency-bin manipulation approaches to all-optical networks---is by nature broad and open to a variety of potential solutions, depending on the characteristics and needs of a given network. So in our view the main contribution of this work lies in formalizing the vision and specifying methods which can be applied quite generally to future designs. In particular, the distinction between the needs of quantum and classical frequency-bin processing steers one toward information-theoretic metrics, rather than quantum-mechanical performance measures, in evaluating networks in the classical context. The explicit numerical solutions illuminate valuable attributes which seem to apply quite generally to AFPs, such as: (i) basic operations require few components (sublinear in $N$) when arbitrary modulation is available; (ii) improved resource provisioning compared to conventional silicon photonic networks could be possible for a sufficiently large number of nodes; and (iii) the interaction between two effects---noise from interchannel crosstalk and reduction in signal levels from scattering---plays a central role in establishing the optimal AFP transformation.

Our primary motivation for the frequency-bin AFP is optical networking in datacenter and HPC environments. The corresponding need to interface with electronic computing nodes thus renders on-chip implementation a necessity, at least for any practical application. However, even beyond such application-based drivers toward miniaturization, on-chip integration offers fascinating opportunities to improve raw \emph{performance} as well. One of the most conspicuous challenges facing frequency-bin AFPs is high insertion loss; for example, in a previous experiment with a tabletop EOM/pulse shaper/EOM cascade, we observed 12.9~dB loss on an input field~\cite{Lu2018a}. On the other hand, an on-chip phase modulator with <0.5~dB loss has already been demonstrated~\cite{Wang2018}, and CMOS-compatible MRR-based pulse shapers with <1~dB loss appear possible from existing foundries~\cite{AIM2019}. Combining both ideal EOM and pulse shaper capabilities on a single chip has yet to be realized, but the rapid trajectory of advances certainly supports confident optimism. Moreover, the physical realization of spectral shaping via dedicated MRRs could markedly reduce the demands placed on high-bandwidth EOMs. In pulse shapers based on spatial dispersion, such as liquid-crystal-on-silicon technology~\cite{Roelens2008}, the minimum frequency spacing is ultimately limited by spectral resolution, setting a hard lower bound on the modulation frequency the EOMs must attain. On the other hand, MRR-based shapers dedicate independent rings to each frequency bin, all of which can be tuned independently~\cite{Wang2015}; thus the frequency spacing can in principle be much smaller, limited only by the MRR linewidth, to prevent crosstalk between adjacent bins. 

Another essential direction for future exploration comprises networking-centric aspects such as the real-time updates an AFP would be expected to implement in a given system. In order to benefit from the reconfigurability possible within the frequency-bin processing paradigm, the AFP must be able to determine the needed functionality (i.e., I/O configuration) in real time and then update the transformation as fast as possible. 
This question of arbitration---managing data traffic flow to prevent errors and packet loss---is generally much more difficult in photonics than electronics, since optical data streams cannot be easily buffered~\cite{Rumley2015}. Thus it will be important to consider adapting specific silicon photonic arbitration protocols~\cite{Vantrease2008, Joshi2009, Pan2009, Li2014} to the AFP paradigm. As one interesting possibility on this front, perhaps ideas from all-optical signal processing~\cite{Willner2014} could be applied for arbitration decisions. For example, all-optical tapped delay lines relying on nonlinear frequency conversion have realized pattern recognition at the channel line rate~\cite{Khaleghi2012}. Conceivably, such an approach could be used to continuously check for AFP update requests, although work remains to assess how these capabilities could be realized with the basic chip-scale components discussed here. Finally, the speed of the physical AFP update, following arbitration, should be extremely high as well. New waveforms can be applied to each EOM as fast as the inverse electro-optic bandwidth, and pulse shapers with $\sim$GHz update speed should be attainable via the use of electro-optic phase shifters, so that with proper engineering, our AFP paradigm should be well-matched to the refresh rates desired on a fast optical network.

\vspace{0.2in}

\section{Conclusion}
\label{conclusion}
We have described an approach for all-optical networking based on wavelength multiplexing and active frequency transformations. Consisting of an alternating chain of temporal phase modulation and line-by-line pulse shaping, our all-optical frequency processor (AFP) is able to realize user-defined and reconfigurable channel mappings designed to connect network nodes operating at different wavelengths. Through numerical simulations, we have explicitly designed one-to-one frequency hoppings and 1-to-$N$ broadcast operations for small network sizes, considering both fully arbitrary and sinewave-only temporal modulation. By introducing a mutual-information-based metric, we reveal how to optimize system design under resource limitations as well. Our results extend the unitary frequency-bin operations of quantum frequency processors to classical signal processing, indicating the value of arbitrary frequency-bin operations in classical as well as quantum networks.

\textit{Acknowledgments.---}We thank N. Lingaraju for discussions. This work was performed in part at Oak Ridge National Laboratory, operated by UT-Battelle for the U.S. Department of Energy under contract no. DE-AC05-00OR22725. Funding was provided by ORNL's Laboratory Directed Research and Development Program.

\

\end{document}